# 2 μm Watt-level Fiber Amplifiers, Lasers, and ASE Sources Pumped by Broadband ASE Pumps

Robert E. Tench, *Senior Member, IEEE*, Jean-Marc Delavaux, *Senior Member, IEEE*, Wiktor Walasik, *Member, IEEE*, Alexandre Amavigan, and Yves Jaouen

*Abstract*— We report the design and demonstration of novel 2 μm band Watt-level fiber amplifiers, fiber lasers, and wideband ASE sources that are pumped with broad spectrum Watt-level ASE sources instead of conventional fiber laser pumps. We show good agreement between the simulations and experimental results for the performance of a single-stage Ho-doped fiber amplifier at 2050nm wavelength pumped by a 50 nm broadband Tm-doped ASE source centered at 1860 nm. Next, we show that the new ASE pumping approach works effectively for a two-stage Ho-doped fiber amplifier as for a single stage Ho-doped fiber amplifier. Then, we demonstrate successfully the pumping of a Tm-doped fiber laser at 2039 nm using an ASE source centered at 1550 nm. Finally, we produce a novel 265 nm wide broadband ASE source at 2 μm by concatenating Tm and Ho ASE sources. Our ASE-based pumping approach is simple and versatile compared to the standard laser based pumping means, and leads to similar device performance.

*Index Terms*—Optical Fiber Devices, Optical Amplifiers, Near-Infrared, Thulium-Doped Fiber Amplifiers, Polarization Maintaining Fiber Amplifiers, 2000 nm.

## I. INTRODUCTION

TWO-micron Holmium-doped fibre (HDF) amplifiers and lasers [1-10] have an important role to play in many space and physics applications which include free-space transmission between earth and space in low atmospheric attenuation wavelength bands (e.g., 2039 nm and 2130 nm) [1,3], generation of high energy 3-5 μm signals using 2.1 μm OPO lasers [4,7], remote molecule sensing and eye-safe LIDAR imaging [5,8,9], and future gravity wave interferometers operating at 2051 nm [10].

HDFs with their broad emission spectrum (from 2 to 2.15 μm) and high optical conversion efficiency ($> 80\%$) are quite attractive for 2 μm sources and amplifiers for these current and future physics and engineering applications(ref). For the HDF amplifiers (HDFAs) and lasers to have a major adoption in these many applications, they need to come in disruptive designs combining several features that include: a good electrical-to-optical conversion efficiency, a high power stability, established reliability in small form-factor packages, and cost-effective architectures.

In this paper we report the design and demonstration of novel 2000 nm band broadband Watt-level HDFAs that are pumped with broad spectrum Watt-level Tm-doped amplified spontaneous emission (ASE) sources instead of conventional semiconductor or fibre laser sources [11]. We also demonstrate the pumping of thulium doped DFB fiber lasers and ASE sources [2] with 1550 nm centered broadband ASE pumps. Our approach is simple and cost effective compared to the standard laser-based pumping means and leads to similar device performance.

Our innovative approach is based on the broadband ASE-optical pumping of the active fiber of the amplifiers or lasers instead of conventional configurations that rely on pumping from semiconductor or fiber laser pumps. Specific laser wavelengths with enough power may be hard to find compared with the straightforward generation of high power ASE sources. In addition, our ASE pumped device alternatives offer the following advantages: simple and versatile pumping design, reduced cost, equivalent optical performance, similar reliability and wall plug efficiency, and compatibility with SWAP for both terrestrial and space-based applications.

Our paper is organized in the following way: Section II presents a simulation model of the Ho-doped fiber adapted to using a broad band ASE source to to calculate the amplifier performance ( i.e., signal gain (G), output power ($P_{out}$), noise figure (NF), optical signal to noise ratio (OSNR) etc. . Section III describes both the configurations of 1) a single stage amplifier and 2) ASE-pump sources centered at 1860nm used in our simulations and experimental results of our HDFA performance. In particular, we report the simulation results for the one stage amplifier for different ASE pump bandwidths. In Section IV we present simulated results for the performance of a two-stage HDFA pumped with ASE shared between both stages, and confirm the validity of our approach. In Section V, we compare our experimental performance evaluation for the single stage ASE pumped HDFA at the 2050nm signal wavelength to our simulations, and obtain overall a good agreement. Next in section VI we demonstrate via experimental results that the ASE pumped approach works for pumping single frequency DFB_FBG Thulium-doped fiber lasers using a broadband ASE source centered at 1550nm [2]. Then we experimentally show the generation of a very large bandwidth ASE source by combining Thulium doped and Holmium doped ASE sources. Section VII is a discussion of our results and the

R. E. Tench, J.-M. Delavaux, W. Walasik, and A. Amavigan are with Cybel LLC, 62 Highland Avenue, Bethlehem, 18017, PA, USA (email: robert.tench@cybel-llc.com, jm@cybel-llc.com).wiktor.walasik@cybel-llc.com, alexandre.amavigan@cybel-llc.com)

Y. Jaouën is with Telecom Paris, Institut Polytechnique de Paris, 19 Place Marguerite Perey, F-91120 Palaiseau, France (email: yves.jaouen@telecom-paris.fr)

significance of our approach. We finish with a summary in Section VIII.

## II. HO MODEL

A 4-level energy transfer model can be considered for optical power generation at 2.1µm range, as shown in Fig. 1(a). [6]. Holmium doped fiber can be pumped at two practical wavelengths: 1150 nm ($^5I_6$ level) [12] and 1850-1950 nm ($^5I_7$ level) [13-15]. Pumping the $^4I_6$ level can be addressed using Ytterbium fiber lasers. However, the large quantum defect between pump and signal reduces the efficiency of pump to signal conversion. Usually the in-band pumping using Thulium fiber lasers is preferred as it offers a large efficiency across the 1.9 – 2.05µm wavelength range.

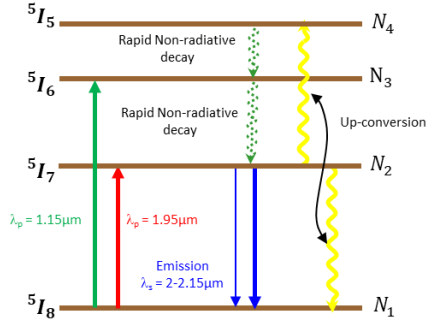

Fig. 1(a). Simplified $Ho^{3+}$ transfer energy level model.

The non-radiative relaxation time of $^5I_5$ and $^5I_6$ levels are very short (i.e. ~1 µs or less [6]) compared to that of $^5I_7$ metastable level (i.e. $\tau_{N_{Ho}} = 0.6$ ms). It results that the corresponding populations $N_3$ and $N_4$ can be neglected compared to populations $N_1$ and $N_2$. Simulations can be performed on a classical 2-level model, similar to EDFA [16], considering the usual set of power propagation and population equations in stationary-state rate regime. The propagation equations (1)—(3) below for $P_p^\pm$, $P_s$ and $P_{ASE}^\pm$ describe the power evolution of pump, signal and ASE along the fiber, where the index λ indicates the spectral dependence for the wavelength dependence for the different powers. The ± indicates the forward (+) and backward (-) propagation. $\alpha_{Ho}$ and $g_{Ho}$ are the absorption and gain coefficient, $l_{core}$ is the core background loss. $\Delta v$ is the resolution bandwidth expressed in frequency domain and $h$ is the Planck constant.

Equation (1) is for the pump wave along the fiber.

$$\frac{dP_{p,\lambda}^\pm}{dz} = \pm\left(\left(\alpha_{Ho,\lambda} + g_{Ho,\lambda}\right)n_{Ho} - \left(\alpha_{Ho,\lambda} + l_{core}\right)\right)P_{p,\lambda}^\pm \quad (1)$$

Equation (2) is for the signal wave along the fiber.

$$\frac{dP_{s,\lambda}}{dz} = \left(\left(\alpha_{Ho,\lambda} + g_{Ho,\lambda}\right)n_{Ho} - \left(\alpha_{Ho,\lambda} + l_{core}\right)\right)P_{s,\lambda} \quad (2)$$

Equation 3 is for the ASE wave along the fiber.

$$\frac{dP_{ASE,\lambda}^\pm}{dz} = \pm\left(\left(\alpha_{Ho,\lambda} + g_{Ho,\lambda}\right)n_{Ho} - \left(\alpha_{Ho,\lambda} + l_{core}\right)\right)P_{ASE,\lambda}^\pm \quad (3)$$
$$\pm 2h\upsilon\Delta\upsilon g_{Ho,\lambda}$$

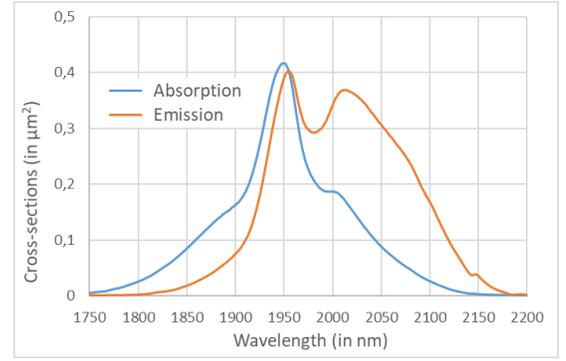

Fig. 1(b). Cross-sections of the $Ho^{3+}$ doped fiber.

The absorption and emission cross-sections are presented in Fig. 1(b). The parameter $n_{Ho}$ is the normalized $Ho^{3+}$ ions population in the excited state respectively, $\zeta_{Ho}$ is the saturation parameter. It is well-known that the up-conversion effect is significant in Holmium doped fiber [17]. Therefore we have introduced 2 different populations, $n_{Ho}^s$ and $n_{Ho}^p$ which represent the relative inversions of single and paired $Ho^{3+}$ ions in the metastable level, similar to EDFA modelling [14]. Tne number equations are given below in (4)—(6).

Equation (4) is for the composition of the full number density.

$$n_{Ho} = \underbrace{n_{Ho}^s}_{\substack{single \\ ions}} + \underbrace{n_{Ho}^p}_{\substack{pairing \\ ions}} \quad (4)$$

Equation (5) is for the number density of single Ho ions in the fiber.

$$n_{Ho}^s = \frac{(1 - x_{ions})\Sigma[\lambda]\,\alpha_{Ho,\lambda}\frac{P_{p,\lambda} + P_{s,\lambda} + P_{ASE,\lambda}}{h\nu_\lambda\zeta_{Ho}}}{1 + \Sigma[\lambda]\,(\alpha_{Ho,\lambda} + g_{Ho,\lambda})\frac{P_{p,\lambda} + P_{s,\lambda} + P_{ASE,\lambda}}{h\nu_\lambda\zeta_{Ho}}} \quad (5)$$

Equation (6) is for the number density of paired Ho ions in the fiber.

$$n_{Ho}^p = \frac{x_{ions}\,\Sigma[\lambda]\,\alpha_{Ho,\lambda}\frac{P_{p,\lambda} + P_{s,\lambda} + P_{ASE,\lambda}}{h\nu_s\zeta_{Ho}}}{1 + \Sigma[\lambda]\,(2\alpha_{Ho,\lambda} + g_{Ho,\lambda})\frac{P_{p,\lambda} + P_{s,\lambda} + P_{ASE,\lambda}}{h\nu_\lambda\zeta_{Ho}}} \quad (6)$$

All the fiber parameters are defined in Table 1 for the Exail Ho-doped fiber used in our simulations and experiments, IXF-HDF-PM-8-125 [1,3,11,13—15]. The propagation equations are solved using the ODE45 solver in Matlab and calculated using a spectral resolution of 0.1 nm for a precise representation of the pump and signal spectra. In this spectral resolution, the broadband pump source consists of multiple appropriately spaced monochromatic components and an overall Gaussian amplitude distribution with a 3 dB width of Δλ and a total integrated power of $P_{pump}$.

| Symbol | Parameter | Value | Unit |
|---|---|---|---|
| $\Phi_{core}$ | Core diameter | 8 | μm |
| $\Gamma$ | Pump/Sig. overlap | 0.76 | |
| $N_{Ho}$ | Ho concentration | $2.75\ 10^{25}$ | m$^{-3}$ |
| $n_{Ho}$ | Normalized pop. inv. | $N_2/N_{Ho}$ | |
| $\sigma_{Ho,a}$ | Ho Cross Sec. Abs. | See Fig. | μm$^2$ |
| $\sigma_{Ho,e}$ | Ho Cross Sec Ems | See Fig. | μm$^2$ |
| $\alpha_{Ho,\lambda}$ | Ho Abs. Coeff. | $N_{Ho}\sigma_{Ho,a}\Gamma$ | m$^{-1}$ |
| $g_{Ho,\lambda}$ | Ho Ems. Coeff. | $N_{Ho}\sigma_{Ho,e}\Gamma$ | m$^{-1}$ |
| $\tau_{N_{Ho}}$ | $N_{Ho}$ Lifetime | 0.6 | ms |
| $A$ | Effective area | $\pi(\Phi_{core}/2)^2$ | μm$^2$ |
| $\zeta_{Ho}$ | Ho Sat. Param. | $A/N_{Ho}\tau_{N_{Ho}}$ | 1/(m.s) |
| $x_{ions}$ | Ion-pairing | 13.5 | % |
| $\lambda_p$ | Pump wavelength | 1860 | nm |
| $\lambda_s$ | Signal Wavelength | [2000-2150] | nm |
| $l_{core}$ | Background loss | 0.2 | dB/m |

Table 1: Fiber parameters used for the simulation

## III. ARCHITECTURE OF THE ASE PUMPED HDFA AND SIMULATION RESULTS

In this section we present an initial simulation study illustrating the performance of a one-stage HDFA pumped by an ASE source. Figure 2 shows the optical configuration of the HDFA under study. The input signal, P$_{in}$, at a wavelength of 2050 nm with a 0 dBm (1 mW) CW power, is coupled into the active Ho-doped fiber F1 (Exail IXF-HDF-PM-8-125) through isolator I1 and wavelength division multiplexer WDM1. A broadband ASE pump light source is coupled into F1 through WDM1 and pumps the Holmium ions in the fiber to produce gain within the emission band (i.e., from 2 to 2.15 μm). The ASE pump has an equivalent Gaussian spectrum centered at λ$_0$ = 1860 nm [11] with a 3-dB width of Δλ and a CW ASE output power of P$_{pump}$. As a result, the signal is amplified along the 2.5-m-long single-clad Ho-doped fiber F1 and exits after isolator I2.

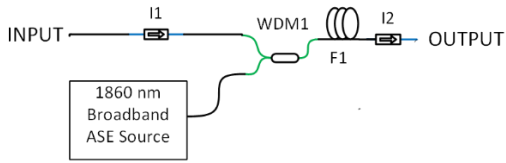

Figure 2. Configuration of a single stage HDFA co-pumped with a broadband ASE source.

Figure 3 shows the architecture of a representative experimental two-stage Tm-doped ASE source (a) and plots the spectral evolution of the source emission as a function of output power P$_{out}$ (b). [2] This graph demonstrates that for a P$_{out}$ up to 1.25 W the ASE source exhibits similar behavior in terms of center wavelength (e.g., 1860 nm) and 3 dB or FWHM emission bandwidth (50 nm), while exhibiting no instability or self-lasing. It should be noted that the 1860 nm center wavelength coincides with one of the pump wavelengths that produces the best power conversion efficiency in the Ho-doped fiber amplifier [13—15].

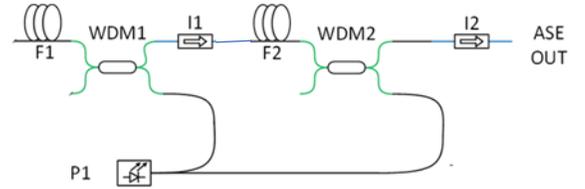

Figure 3(a). Architecture of experimental two-stage Tm-doped ASE source

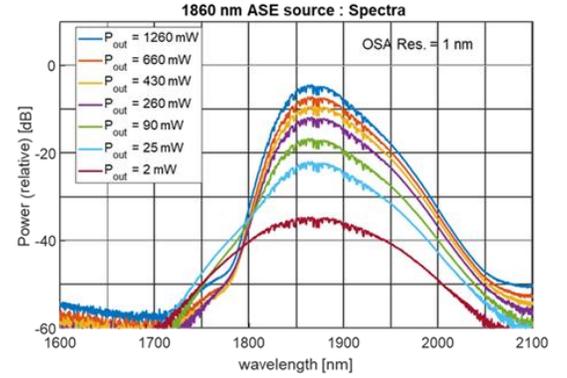

Figure 3(b). Measured output spectra vs. CW ASE output power.

In using the 1860 nm ASE source as a pump for our single stage HDFA in Figure 1, we have simulated the P$_{out}$ and noise figure performance for 1 mW input signal power P$_{in}$ at 2050 nm wavelength [15]. In this plot the signal input and output powers are measured at the input and output of F1. The 3 dB width of the ASE pump was set to 50 nm to match the experimental 3 dB spectral width, and a monochromatic pump source with a negligible linewidth (i.e., < 0.1 nm) was also plotted for comparison. The simulation results in Figure 4 for the 50 nm (3 dB) ASE source indicate that 1) the P$_{out}$ increases linearly with pump power, with a P$_{out}$ of 0.9 W for a 2 W pump and a 38% slope efficiency, and 2) the NF value plateaus around 3.7 dB for more than 0.5 W of 1860 nm pump power. In contrast, with the monochromatic source, there is no change in the NF evolution and an improvement of P$_{out}$ conversion efficiency to 40%. In other terms, the ASE pump produces similar performance to the monochromatic laser pump.

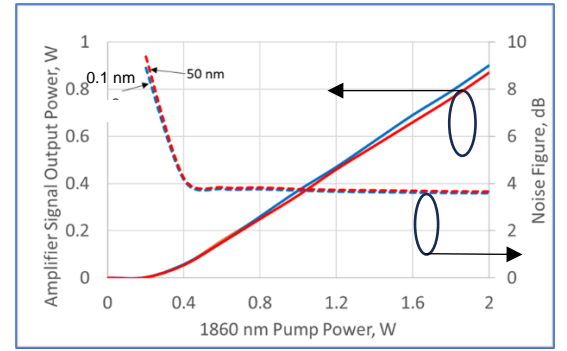

Figure 4. HDFA P$_{out}$ and NF vs. ASE pump power and Δλ. The blue curves are for Δλ = 0.1 nm, and the red curves are for Δλ = 50 nm.

In the two stage ASE source configuration, the 3 dB width of the spectral emission can be selected by inserting a mid-stage bandpass filter centered at a wavelength within the natural ASE emission spectrum. Basically, the first stage seeds and also saturates the second stage Tm-doped fiber amplifier to extract maximum ASE output power. As an example, we have simulated the performance of a two-stage ASE source with a bandpass filter centered at 1860 nm with variable spectral widths from 0.5 nm to 50 nm as shown in Figure 5. These simulations were carried out with proprietary code from Cybel written in Microsoft Visual Basic 5.0.

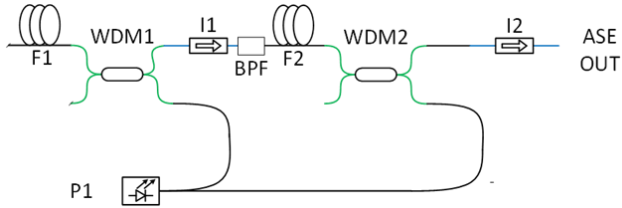

Figure 5. Two stage 1860 nm ASE source with interstage bandpass filter.

The corresponding evolution of the ASE output power as a function of wavelength and spectral width is plotted in Figure 6. As the filter width decreases, the peak power increases while the total power remains almost the same. This design enables us to concentrate the ASE power over a limited bandwidth and thus optimize the pumping efficiency of the HDFA. In this plot we note that the total or CW power of the ASE source is constant at 2.0 W for all the spectral widths investigated, from 0.5 nm to 50 nm.

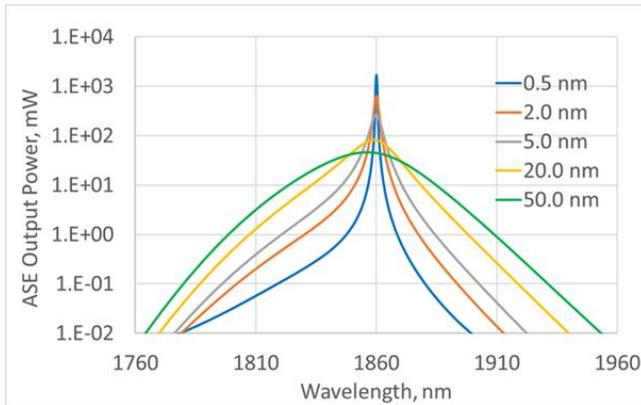

Figure 6. Simulated output spectra for a 2 W average output power TDF ASE source, as a function of 3 dB BPF bandwidth. Resolution bandwidth is 1 nm, center wavelength of filter = 1860 nm.

To establish the effect of spectral bandwidth on the output power of the HDFA, we conducted simulations [11, 13—15] on the Ho-doped fiber amplifier in Figure 1 by setting the pump power to 1.4 W and changing the spectral width from 0.5 nm to 80 nm. Here the input and output signal powers are measured at the input and output of the active Ho-doped fiber. Figure 7 plots the simulated results of $P_{out}$ as a function of the spectral width for $P_{pump}$ = 1.4 W. For reference purposes the output power using a monochromatic pump source is plotted with a red star on the left-hand axis. The broadband ASE source is just as efficient as the monochromatic laser source for spectral widths $\Delta\lambda$ below 20 nm. These results confirm the effectiveness of using broadband ASE pump sources for Ho-doped fiber amplifiers.

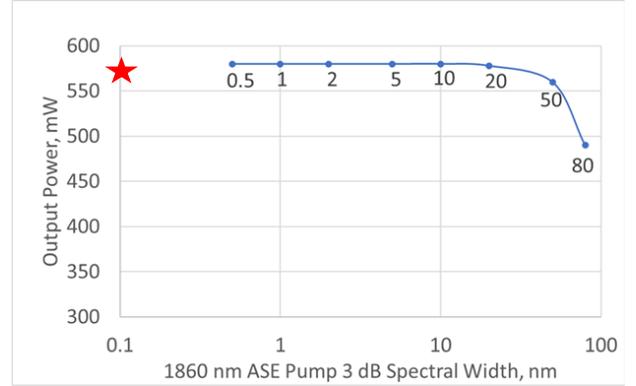

Figure 7. Simulated signal output power at 2050 nm as a function of pump spectral width $\Delta\lambda$ (blue data points).

Figure 8 shows the simulated output spectrum for the HDFA under study (Figure 1), for a pump 3 dB spectral width of 10 nm and a total ASE pump power of 1.4 W [11,13—15]. The broad spectral nature of the ASE pump source is clearly displayed in the 1860 nm region of the spectrum. The amplified output signal portion of the spectrum from 1950 nm to 2200 nm is identical to the spectrum obtained with a monochromatic pump source. The optical signal to noise ratio (OSNR) is 65 dB in a resolution of 0.1 nm. For a counter pump single stage amplification, the residual pump will not be present in the amplifier output.

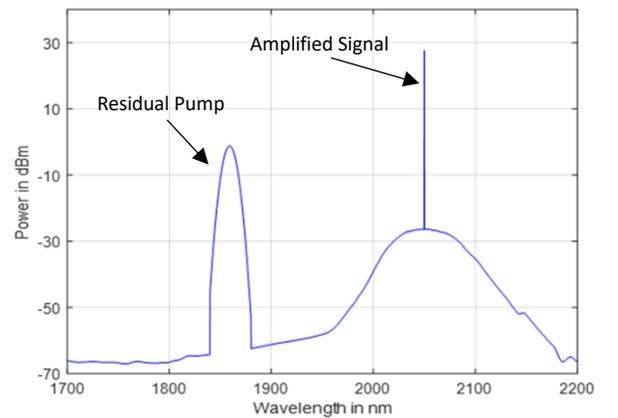

Figure 8. Simulated output spectrum of the co-pumped HDFA for ASE $\Delta\lambda$ = 10 nm, $P_{in}$ = 0 dBm at 2050 nm, and RBW = 1.0 nm.

## IV. SIMULATIONS OF TWO STAGE HDFA WITH BROADBAND ASE PUMPING

In this section we present simulations of a two stage HDFA with broadband 1860 nm ASE pumping of both stages, to demonstrate the effectiveness of broadband pumping for multi-stage fiber amplifier architectures.

The optical schematic diagram of our two stage ASE pumped HDFA is shown in Figure 9.

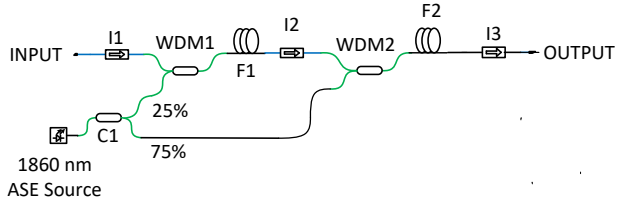

Figure 9. Optical Architecture of a Two-stage HDFA with Broadband 1860 nm ASE Pumping

Here the Ho-doped fiber is Exail/iXblue IXF-HDF-PM-8-125. The fiber lengths for our simulations are F1 = 2.5 m and F2 = 1.5 m. The broadband 1860 nm pump powers are 1.6 W for stage 1 and 5.0 W for stage 2 out of coupler C1 for a total pump power of 6.6 W at the 1860 nm ASE pump source. The input signal is 0 dBm (1 mW) at 2050 nm. Insertion losses of the optical isolators I1-I3 are 1.2 dB, and the pump and signal losses in the WDM couplers WDM1 and WDM2 are 0.6 dB.

Figure 10 shows the simulated fiber coupled signal output power at 2050 nm as a function of the width of the 1860 nm ASE pump source. Here the width of the pump source is varied from 1 nm to 80 nm, and the red star on the left axis is the signal output power for a monochromatic pump source. We see that the output power of the amplifier remains within 0.2 dB of its monochromatic value for filter widths of 1-50 nm, verifying the effective operation of broadband ASE pumping for a two stage amplifier as well as for one stage amplifiers. Figure 11 shows the simulated output spectrum for an 1860 nm broadband pump width of 10 nm.

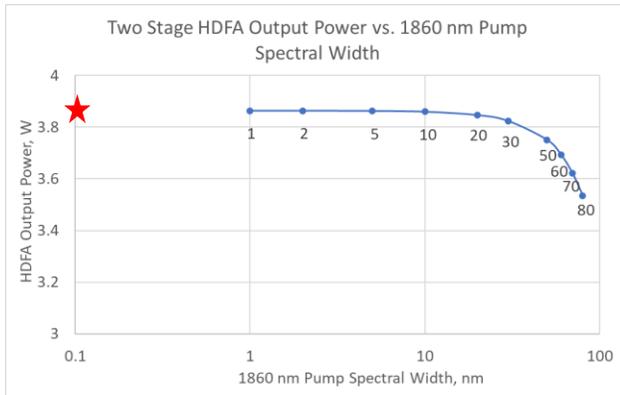

Figure 10. Simulated 2050 nm Signal Output Power for the Two Stage HDFA as a Function of Pump Spectral Width

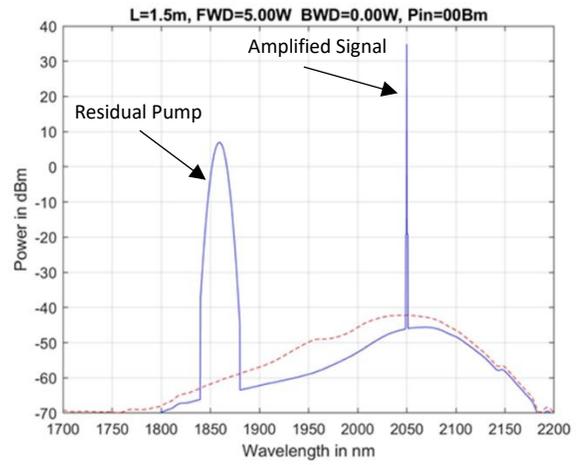

Figure 11. Output Spectrum (blue) for the Two Stage HDFA with an 1860 nm Broadband Pump Spectral Width of 10 nm. Backward ASE power is shown in red.

## V. EXPERIMENTAL RESULTS FOR AN ASE-PUMPED HDFA

Experimental results for the single stage amplifier configuration in Figure 1 are plotted in Figure 12, which shows signal output power vs. pump power for the single stage HDFA using both a broadband ASE source centered at 1860 nm (FWHM = 60 nm) and a narrowband fiber laser source (i.e, 0.1nm) at 1860 nm. In this plot the signal input and output powers are measured at the input and output pigtail fibers of the amplifier, with a corresponding output signal coupling loss relative to the output of the active fiber F1 of 3.2 dB. We observe that the output power performance of the HDFA is similar when pumped with both a broadband ASE source and a narrow band fiber laser source, with a difference of -0.5 dB between the ASE pumped amplifier and the fiber laser pumped amplifier. This difference in output power is entirely consistent with the simulations in Figure 7, and confirms the predictions that ASE pumping is equally as effective as single frequency fiber laser pumping. The experimental noise figure for 0 dBm input power is $4.2 \pm 1$ dB which is consistent with simulations in Figure 3.

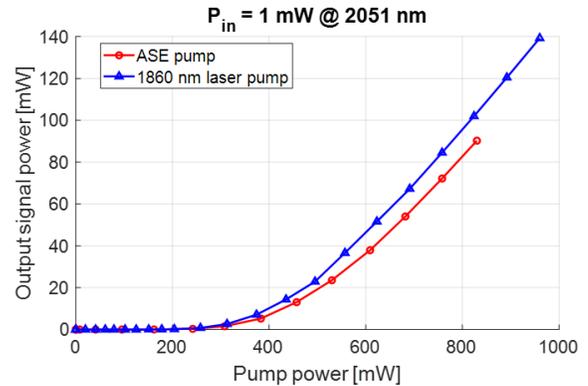

Figure 12. Experimental $P_{out}$ vs. $P_{pump}$ for ASE pumping and fiber laser pumping.

Figure 13(a) plots the simulated and experimental output power for the HDFA with fiber laser pumping at 1860 nm, and Figure 13(b) shows the simulated and experimental output power for the HDFA with ASE source pumping at 1860 nm. Both plots show good agreement between simulation and experiment, verifying our approach both in simulations and in experiment to this novel ASE pumped HDFA design.

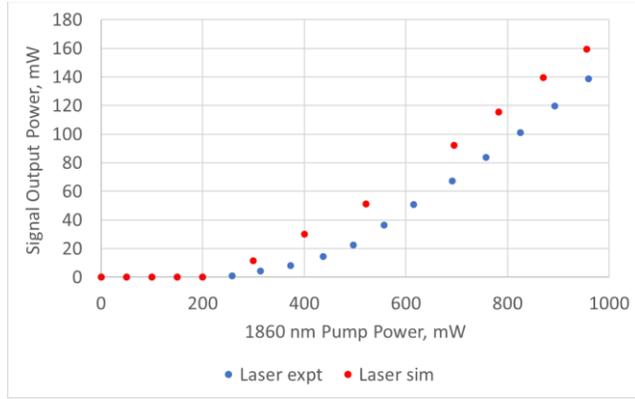

Figure 13(a). Simulated and experimental signal output power for the HDFA with fiber laser pumping at 1860 nm.

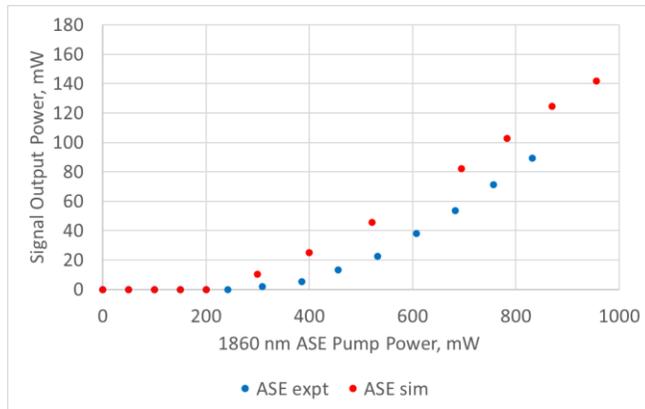

Figure 13(b). Simulated and experimental HDFA signal output power as a function of ASE pump power at 1860 nm

Figure 14 plots the output spectra for the HDFA with both broadband 1860 nm ASE pumping (blue) and single frequency 1860 nm fiber laser pumping (red). We see that the output powers and HDFA ASE output within the signal band of 2000–2100 nm are essentially identical as expected, while the pump band feedthrough in the 1800–2000 nm band is consistent with expectations.

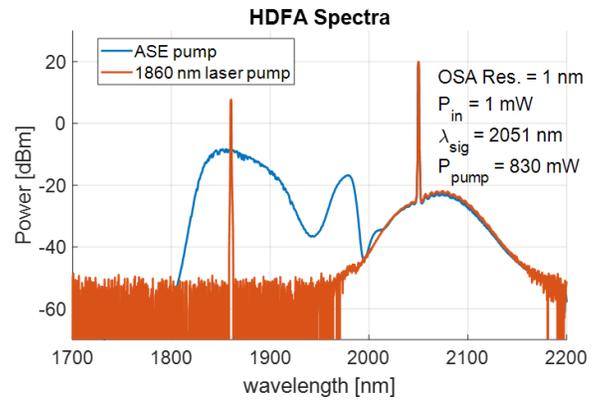

Figure 14. Experimental output spectra for 1860 nm broadband ASE pumping and 1860 nm single frequency fiber laser pumping.

VI. EXPERIMENTAL RESULTS FOR ASE-PUMPED THULIUM DOPED DFB FIBER LASER AND BROADBAND ASE SOURCE

Next, to illustrate the applicability of an ASE source to pump fiber lasers, we have characterized the performance of a 2039 nm PM Tm-doped DFB FBG laser [2]. The system under consideration is shown in the inset of Fig. 15 (left). The DFB FBG laser is pumped with an unseeded Er/Yb-doped fiber amplifier (EYDFA) which acts as a 1.5 µm ASE source. The performance of the ASE pumped laser was contrasted with results obtained using a master oscillator power amplifier (MOPA) pump at three different wavelengths (i.e., 1550, 1560, and 1565 nm) in order to change the absorption coefficient in the fiber laser. The MOPA consisted of a narrow- linewidth tunable laser (Santec TSL-710) amplified by the 1W EYDFA. The pump spectra are shown in the main panel of Fig. 15. The right panel of Fig. 16 shows the signal output power of the DFB FBG laser as a function of pump power for different pump types. We observe that pumping with a broad-band ASE is only 8% less efficient than using a monochromatic pump at 1550 nm. We note that by placing a few-nm-wide ASE filter between the two stages of the EYDFA, the ASE pumping approach enables the selection of both the ASE center wavelength and its bandwidth [11], leading to an increased optical efficiency and a reduced low-frequency noise contribution to the DFB FBG laser emission.

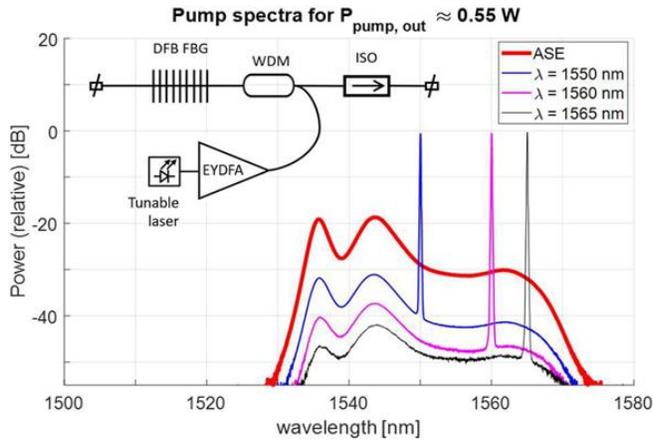

Figure 15. Spectra of 1.5 um ASE and MOPA pumps. The inset shows the schematic of the 2039 nm PM FGB DFB architecture.

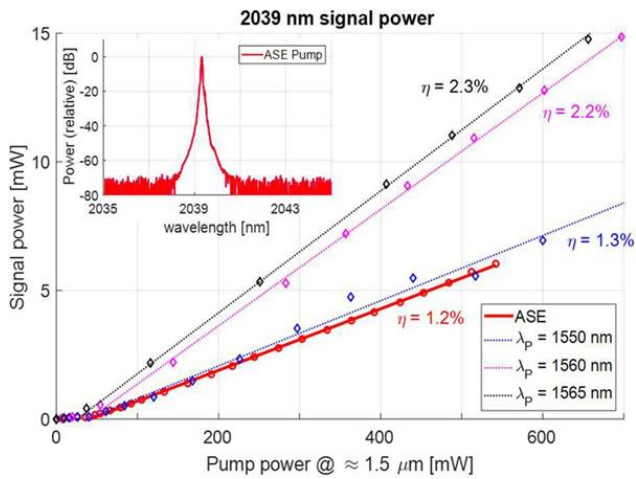

Figure 16. Power emitted by the laser as a function of pump power, for four different pumps. Inset shows the spectrum of the laser when pumped with an ASE source.

Finally, we show that using a broadband ASE to pump a rare-earth-doped ASE source covering a longer wavelength range leads to generation of an ultra-wide 2000 nm band ASE source [2]. In particular, we used a Tm-based ASE source to pump a length of Ho-doped fiber, as shown in diagram (A) in Fig. 17. The output spectra at different integrated output power levels are shown in Fig. 18. We observe that for output powers above 10 mW, the 20-dB bandwidth of this ASE source is 265 nm, spanning from 1825 nm to 2090 nm. To the best of our knowledge, this is the broadest 2000 nm band ASE source delivering more than 50 mW of integrated output power (see [18—22]). Further simplification of the ASE architecture and equalization of the spectral shape are possible by modifying the system topology (as shown schematically in (B) in Fig. 17), optimizing the fiber lengths, and controlling the ASE pump polarization.

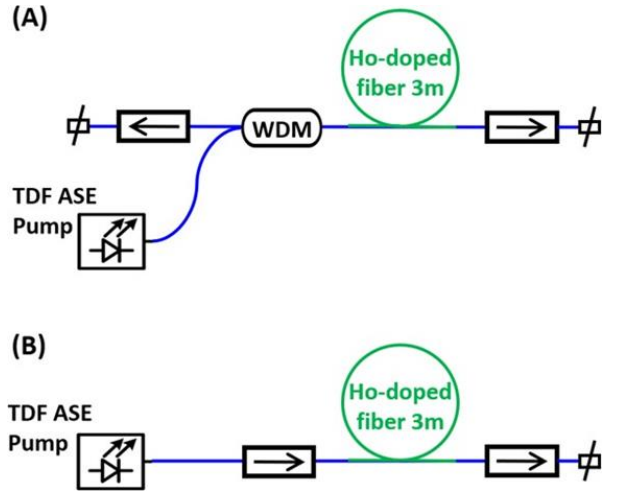

Figure 17. (A) Ultra-broad ASE source with spectrum shown in Fig. 13 at different ASE pump power levels. Configuration (B) shows a simplified broadband ASE topology.

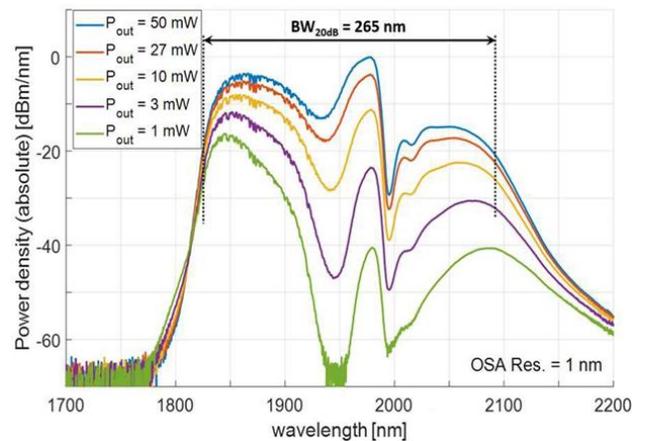

Figure 18. Output spectra of ultra-broad ASE source (A) in Fig. 17 at different output power levels. Source width = 265 nm at 20 dB down for $P_{out}$ = 50 mW.

## VII. DISCUSSION

In our study of architecture and performance of ASE pumped HDFAs in Sections III (single stage amplifiers) and IV (multistage amplifiers), we demonstrated that using spectrally shaped narrowband and broadband ASE sources as pumps for Ho-doped fiber amplifiers is a simple, straightforward, efficient, and technically effective means of building high performance amplifiers for many venues, including but not limited to terrestrial and space-based applications. Pumping with properly configured ASE sources was found to be equivalent in optical-optical conversion efficiency to pumping with narrow linewidth semiconductor or fiber laser sources.

We further note that ASE sources are not expected to exhibit the relaxation oscillations, and thus amplitude noise or RIN, found in fiber laser pump sources. We therefore expect the RIN of ASE-pumped HDFAs and other fiber amplifiers and lasers to be significantly reduced in comparison to the current generation of fiber amplifiers pumped by fiber laser sources. This point will be discussed in detail in future publications.

In Section V, we outlined the experimental performance of ASE pumped single stage HDFAs and showed good agreement between our simulations and our experimental results.

In Section VI, we presented the experimental performance 2000 nm band single frequency Thulium doped fiber lasers pumped with 1550nm centered broadband ASE sources. We showed highly successful operation of both types of advanced photonics sources, and outlined the ways in which the broadband ASE pumping of single frequency DFB-FBG fiber lasers could be made even more efficient than demonstrate in Figures 14 and 15. Finally we demonstrate the largest bandwidth ASE source by concatenating Thulium- and Holmium-doped ASE sources.

Overall, our results show the clear advantages of using ASE sources as an alternative and versatile means for pumping fiber amplifier, laser or ASE sources with performance comparable to that obtained with traditional laser pumps.

## VIII. SUMMARY

We have reported the design of novel 2000 nm band Watt-level fiber amplifiers, lasers, and wideband ASE sources that are pumped with broad spectrum Watt-level ASE sources instead of conventional narrow linewidth semiconductor or fibre laser sources. We have demonstrated through simulations, and experiments the success of this novel pumping approach. Our approach is simple and cost effective than the standard laser based pumping means, and leads to similar amplifier performance. We expect to see many immediate and significant applications of our novel ASE-based pumping of fiber amplifiers, fiber lasers, and fiber-based ASE sources not only for Ho-doped hosts but also for Yb-, Er-, Er-Yb, Tm, Ho-Tm, and Bi-doped fibers as well.